\documentclass[10pt]{iopart}

\usepackage{iopams}  
\usepackage{graphicx}
\usepackage[breaklinks=true,colorlinks=true,linkcolor=blue,urlcolor=blue,citecolor=blue]{hyperref}
\usepackage{epstopdf}
\usepackage{caption}

\begin{document}
\title[]{Application of bond valence method in the non-isovalent semiconductor alloy (GaN)$_{\rm{1-x}}$(ZnO)$_{\rm{x}}$}

\author{Jian Liu}
\address{Department of Physics and Astronomy, Stony Brook University, Stony Brook, NY 11794-3800, United States.}
\eads{\mailto{Jian.Liu@stonybrook.edu}}

\begin{abstract}
This paper studies the bond valence method (BVM) and its application in the non-isovalent semiconductor alloy (GaN)$_{\rm{1-x}}$(ZnO)$_{\rm{x}}$. Particular attention is paid to the role of short-range order (SRO). A physical interpretation based on atomic orbital interaction is proposed and examined by density-functional theory (DFT) calculations. Combining BVM with Monte-Carlo simulations and a DFT-based cluster expansion model, bond-length distributions and bond-angle variations are predicted. The correlation between bond valence and bond stiffness is also revealed. Finally the concept of bond valence is extended into the modelling of an atomistic potential.
\end{abstract}


\maketitle

\ioptwocol

\section{Introduction}
Isovalent semiconductor alloys are extensively studied since their structural, electronic and optical properties can be tuned by varying the alloy composition. More recently, non-isovalent semiconductor alloys begin to attract attention. For example, the pseudobinary (GaN)$_{\rm{1-x}}$(ZnO)$_{\rm{x}}$ alloy is attractive for its high efficiency in photocatalytic water splitting\cite{Maeda1}. From the theoretical perspective, my co-workers and I have recently predicted strong short-range order (SRO) in the (GaN)$_{\rm{1-x}}$(ZnO)$_{\rm{x}}$ alloy due to its non-isovalent nature\cite{paper1}. The role of SRO in determining the atomic, electronic and vibrational properties is also revealed. To fulfill the local charge neutrality, the substitutional SRO is accompanied with and compensated by atomic deviation from the ideal lattice positions. Therefore it is imperative to study the composition-, temperature- and SRO-dependent $(x,T,\Pi)$ structural relaxations. In Ref. \cite{paper1} the (GaN)$_{\rm{1-x}}$(ZnO)$_{\rm{x}}$ alloy is efficiently represented by a SRO-modified version of the Special Quasirandom Structure (SQS)\cite{SQS} approach. The SQS-based approach ``mimics'' the statistics of the correlations. The present study aims directly at statistically reliable predictions of bond-length distribution and bond-angle variation.
\vspace{1pc}
\newline
The bond valence method (BVM) is widely adopted in solid state chemistry for various applications including
prediction of molecular geometry\cite{Brown_review}, construction of atomic potentials for perovskite oxides\cite{perovskite1}\cite{perovskite2}, and calculation of the acidity constant p$K_a$\cite{MUSIC}\cite{pKa}. Its power in predicting the energetics for non-isovalent semiconductor alloys is recently demonstrated\cite{WangPRL}\cite{WangPRB}. In inorganic chemistry the BVM is commonly recognized as an empirical tool, the underlying physics of which is not widely discussed. For example, the fact that bond valence correlates strongly with bond length\cite{Brown_review} reflects the connection between bond valence and bond-length-dependent transferable force constant\cite{Ceder}. Also the correlation between total energy and bond valence\cite{Brown_review} is not yet fully understood. Brown\cite{Brown_review} proposed a ``more rigorous but less physical'' analogy of the Kirchhoff circuit law which treated the bond valence network as a capacitive electric circuit. Burdett\cite{orbital} interpretated BVM from the molecular orbital basis. There is also some similarity between the bond valence and the Mayer bond order\cite{Mayer}. In the present study, particular attention is paid to the theoretical standing of the BVM. Density-Functional Theory (DFT) calculations are performed wherever necessary.
\vspace{1pc}
\newline
The present study also has a computational motivation. Electronic structure methods can handle fairly large supercells (e.g., over 100 atoms). For non-isovalent semiconductor alloys, a large supercell is favored in order to average out the fluctuation error due to the finite size. As large structural relaxations are expected, accurate DFT total energy and force calculations are computationally expensive. Therefore it is desirable to pre-relax the internal atomic positions in an economical way. The BVM-predicted bond lengths and bond angles suit this purpose well.

\section{The Bond Valence Method}
The BVM is extensively discussed in Ref. \cite{Brown_book}. Each nearest-neighbor cation-anion bond is assigned a bond valence $v_{IJ}$. Next nearest-neighbor cation-cation/anion-anion interactions are neglected. The bond valence sum (BVS) of an atom is defined as the sum of the bond valences surrounding the atom. Each atom has an ionic valence $V$ equal to its corresponding formal ionic charge. By convention, $V$(Ga)=$+$3, $V$(N)=$-$3, $V$(Zn)=$+$2, $V$(O)=$-$2, $v_{IJ}=-v_{JI}$. Of crucial importance for non-isovalent semiconductor alloys are two rules: (1) the $valence$ $sum$ $rule$ $V(I)=\sum_Jv_{IJ}$, and (2) the $valence$ $loop$ $rule$ $\sum_{loop}v_{IJ}=0$. The $valence$ $sum$ $rule$ is an equivalent statement of the principle of local charge neutrality, with the correlation $P_{IJ}\propto v_{IJ}$ where $P_{IJ}$ is the Mulliken overlap population\cite{Mulliken1}\cite{Mulliken2}. The $valence$ $loop$ $rule$ is also known as the $equal$ $valence$ $rule$, since the zero circulation condition is equivalent to the minimization of $\sum_{I,J}v_{IJ}^2$ (see for example the appendix in Ref. \cite{WangPRB}). The solution is a set of \{$v_{IJ}$\} which minimizes the measure of the total energy $E=\alpha\sum_{I,J}v_{IJ}^2$ ($\alpha$ is the correlation constant) under the constraint of the $valence$ $sum$ $rule$.
\vspace{1pc}
\newline
As for the measure of the total energy, in solid state language, a perturbation expansion of the orbital interaction energy reads
\begin{equation}
E=\sum_{i\in I}\epsilon_i=\sum_{i\in I}\left(\epsilon_i^0+\sum_{j\in J}\frac{|\langle\phi_i^0|H'|\phi_j^0\rangle|^2}{\epsilon_i^0-\epsilon_j^0}\right)
\end{equation}
where $\epsilon_i^0$ and $\phi_i^0$ denote atomic energy and orbital respectively. Capital $I,J$ and lowercase $i,j$ refer to atomic and orbital indices respectively. Assuming the correlation $H_{ij}\propto S_{ij}$ ($H_{ij}$ is the matrix element $\langle\phi_i^0|H'|\phi_j^0\rangle$ and $S_{ij}$ is the overlap integral $\langle\phi_i^0|\phi_j^0\rangle$), the relaxation energy $E-\sum\epsilon_i$ is then approximately equal to $\alpha_{IJ}\sum_{i\in I, j\in J}S_{ij}^2$, where the denominator $\epsilon_i^0-\epsilon_j^0$ is absorbed into a correlation constant $\alpha_{IJ}$. The overlap integral $S_{ij}(r,\theta,\phi)$ can be expressed in a separable form $S_{ij}(r)f(\theta,\phi)$\cite{orbital}. The angular dependence is lifted after summing $\sum_{i\in I, j\in J}f^2(\theta,\phi)$ over all the interacting orbital pairs. The summation over orbital pairs then reduces to the summation over atom pairs. Finally the measure of the total energy $E=\alpha\sum_{I,J}v_{IJ}^2$ is obtained, with the assumptions $v_{IJ}\sim S_{IJ}\equiv\sqrt{\sum_{i\in I, j\in J}S_{ij}^2}$ and $\alpha_{IJ}=\alpha$, while $\alpha$ is to be fitted by DFT total energy calculations. In the analogy of the Kirchhoff circuit law, the bond capacitances are all equal\cite{Brown_review}, which is equivalent to assuming $\alpha$ equal for different types of atomic pairs.
\vspace{1pc}
\newline
The radial dependence of $S_{ij}(r)$ leads naturally to the empirical exponential correlation between bond valence and bond length 
\begin{equation}
v_{IJ}=\exp\left(\left(R_{IJ}^0-R_{IJ}\right)/b_{IJ}\right)
\end{equation}
where $R_{IJ}$ is the observed bond length while $R_{IJ}^0$ and $b_{IJ}$ are empirically fitted bond valence parameters for $I$-$J$ bond. $b_{IJ}$ measures the bond softness and is usually taken as a universal constant of $0.37\AA$, while $R_{IJ}^0$ is experimentally determined from structural data of related materials\cite{BVM_para}. In the present study, the disordered alloy offers abundant structural data. Therefore $R_{IJ}^0$ and $b_{IJ}$ are fitted to DFT calculations instead. The bond-angle variation depends on the higher-order terms of orbital interactions in the perturbation expansion. In general the bond bending force is weaker than the bond stretching force. In the present study, an empirical relation\cite{Brown_book} is used for the crude prediction of anion-cation-anion angles
\begin{equation}
\theta_{ICJ}=109.5+k(v_{CI}+v_{CJ}-V_C/2)
\end{equation}
where $k$ is an empirical constant (equal to 15.3$^\circ$ per valence unit (v.u.) in Ref. \cite{Brown_book}), $v_{CI}$ and $v_{CJ}$ are the bond valences of the two ligand bonds, and $V_C$ is the ionic valence of the central cation. Finally, taking into account the constraints of bond lengths and bond angles, the tetrahedrally coordinated alloy lattice is over-constrained. A cost function can be assigned to the constraints in order to perform the pre-relaxation.
\section{Computational Methodology}
Here a brief outline is given of the computational methods. Details are in Ref. \cite{paper1}. An Ising-type model Hamiltonian for the (GaN)$_{\rm{1-x}}$(ZnO)$_{\rm{x}}$ alloy was constructed by Li\cite{LL} using a DFT-based cluster expansion method\cite{CE1}\cite{CE2}\cite{CE3}. Monte-Carlo simulations are then performed on the constructed cluster expansion model using the $ATAT$ package\cite{ATAT1}\cite{ATAT2}\cite{ATAT3}. For each $(x,T)$ of interest, a thermodynamic ensemble of configurations is generated. For each configuration the bond valences can then be determined using BVM. At this stage only the site occupancies are needed. One could in principle minimize the measure of the total energy $E=\alpha\sum_{I,J}v_{IJ}^2$ with respect to the set of bond valences \{$v_{IJ}$\}. Unlike the Ising-type cluster expansion model, BVM model is essentially short-ranged since the set of bond valences $\{v_{IJ}\}$ forms a interactive network. An iterative scheme\cite{iterative} is used to apply the $equal$ $valence$ $rule$, which generally yields better computational efficiency. Finally the bond-length distribution and bond-angle variation are obtained using the empirical correlations introduced earlier in this section.
\vspace{1pc}
\newline
For most of the results presented in this paper, the Perdew-Burke-Ernzerhof (PBE)\cite{PBE} version of the exchange-correlation functional is used. Kohn-Sham wavefunctions are expanded in a variationally optimized double-$\zeta$ polarized (DZP) basis set, as implemented in the {\sc SIESTA} package\cite{SIESTA}. Ga-3d and Zn-3d electrons are treated explicitly as valence electrons. The $k$-point mesh is chosen to be equivalent to a $6\times6\times4$ mesh for the 4-atom wurtzite unit cell. Pseudopotentials for all the atomic species are available from the {\sc SIESTA} homepage\cite{homepage}. DFT calculations are performed for two reasons: (1) The correlations $P_{IJ}\propto v_{IJ}$, $H_{ij}\propto S_{ij}$ and $v_{IJ}\sim S_{IJ}$ are crucial for the interpretation of BVM and are therefore examined first; (2) The BVM parameters are to be fitted to DFT calculations, after which bond-length distribution and bond-angle variation can be predicted by BVM. I construct three representative 432-atom supercells at $x=0.25$, $0.5$ and $0.75$ for the former purpose, and use a thermodynamic ensemble equilibrated at the experimental synthesis temperature $T=1,123$K\cite{Maeda1} with 72-atom supercells for the latter purpose.

\section{Results and Discussions}
\subsection{Examination of BVM}
In Fig. 1 the correlations $P_{IJ}\propto v_{IJ}$ for different types of bonds are shown. One should keep in mind that the Mulliken population $P_{IJ}$ has no strict physical sense due to its sensitivity to the atomic basis set used in the projection. Therefore in present study, only the qualitative correlation is discussed. The correlation $H_{ij}\propto S_{ij}$ is in reality adopted in the extended H\"uckel method\cite{Huckel} where the off-diagonal Hamiltonian matrix elements $H_{ij}$ are approximated by the corresponding diagonal Hamiltonian matrix elements and the overlap integral through $H_{ij}=KS_{ij}(H_{ii}+H_{jj})/2$. In Fig. 2 the correlations $H_{ij}\propto S_{ij}$ between the first $\zeta$ numerical atomic orbitals of different species are shown. Since Ga and O are more electronegative than Zn and N respectively, Ga-4s and O-2p have deeper atomic energy levels than Zn-4s and N-2p. This explains why the Ga-O curve lies higher than the Zn-N curve. In Fig. 3 the correlations $v_{IJ}\sim S_{IJ}$ for different types of bonds are shown. The linearity of the correlations validates the interpretation of BVM proposed in the present study.
\begin{figure}[htb!]
\includegraphics[scale=0.28]{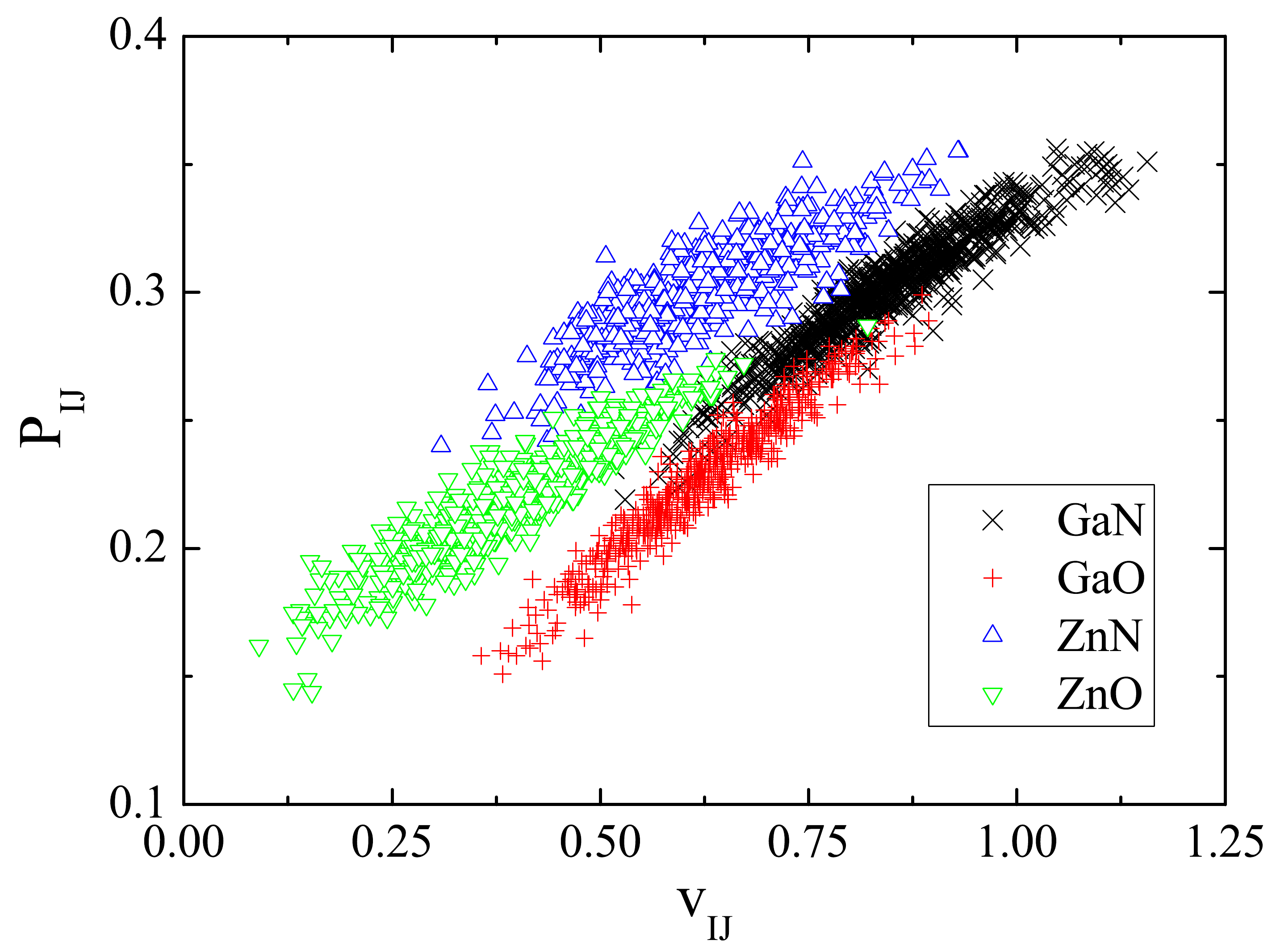}
\caption{The correlation $P_{IJ}\propto v_{IJ}$ for different types of bonds.}
\end{figure}
\begin{figure}[htb!]
\includegraphics[scale=0.28]{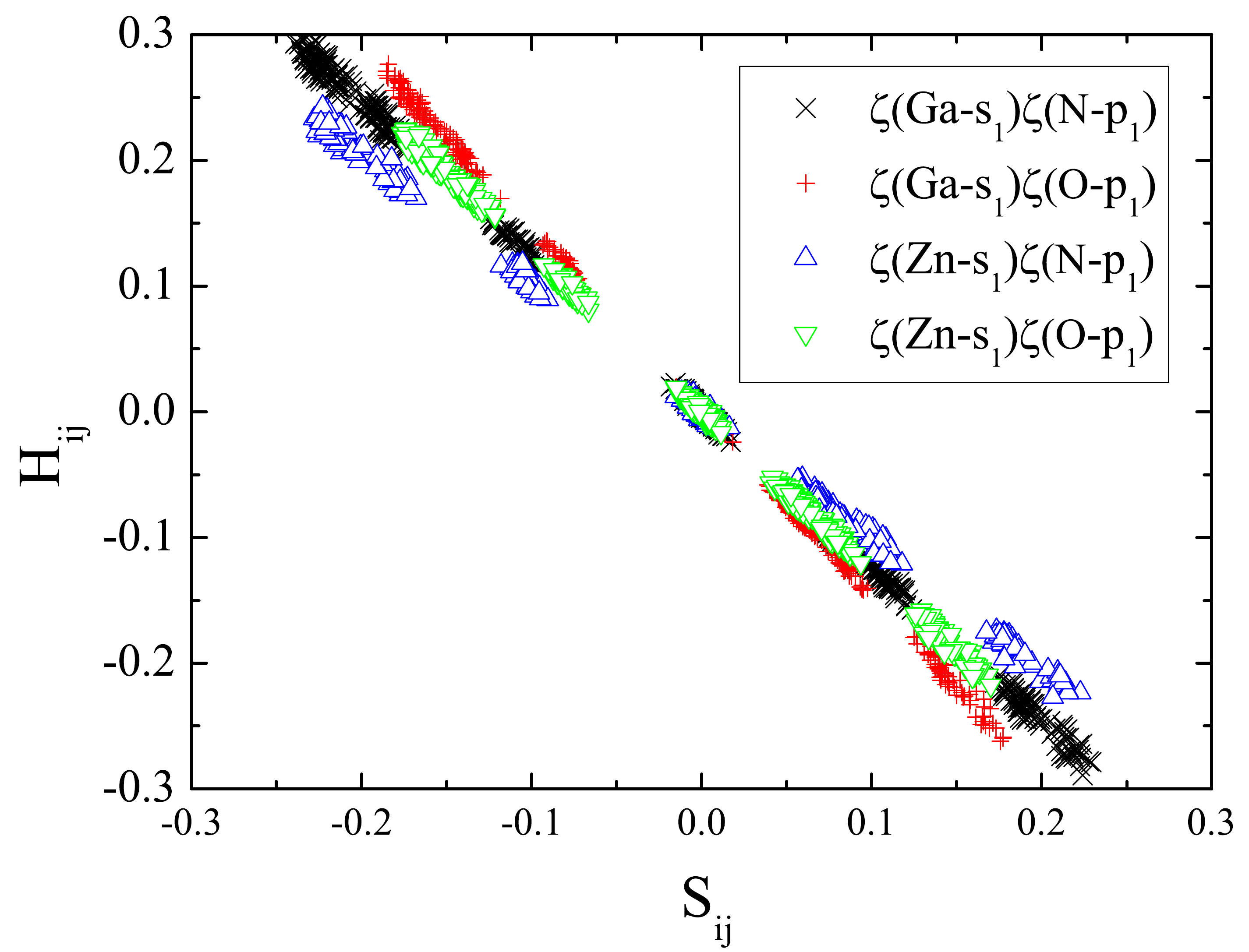}
\caption{The correlation $H_{ij}\propto S_{ij}$ between the first $\zeta$ numerical atomic orbitals of different species.}
\end{figure}
\begin{figure}[htb!]
\includegraphics[scale=0.28]{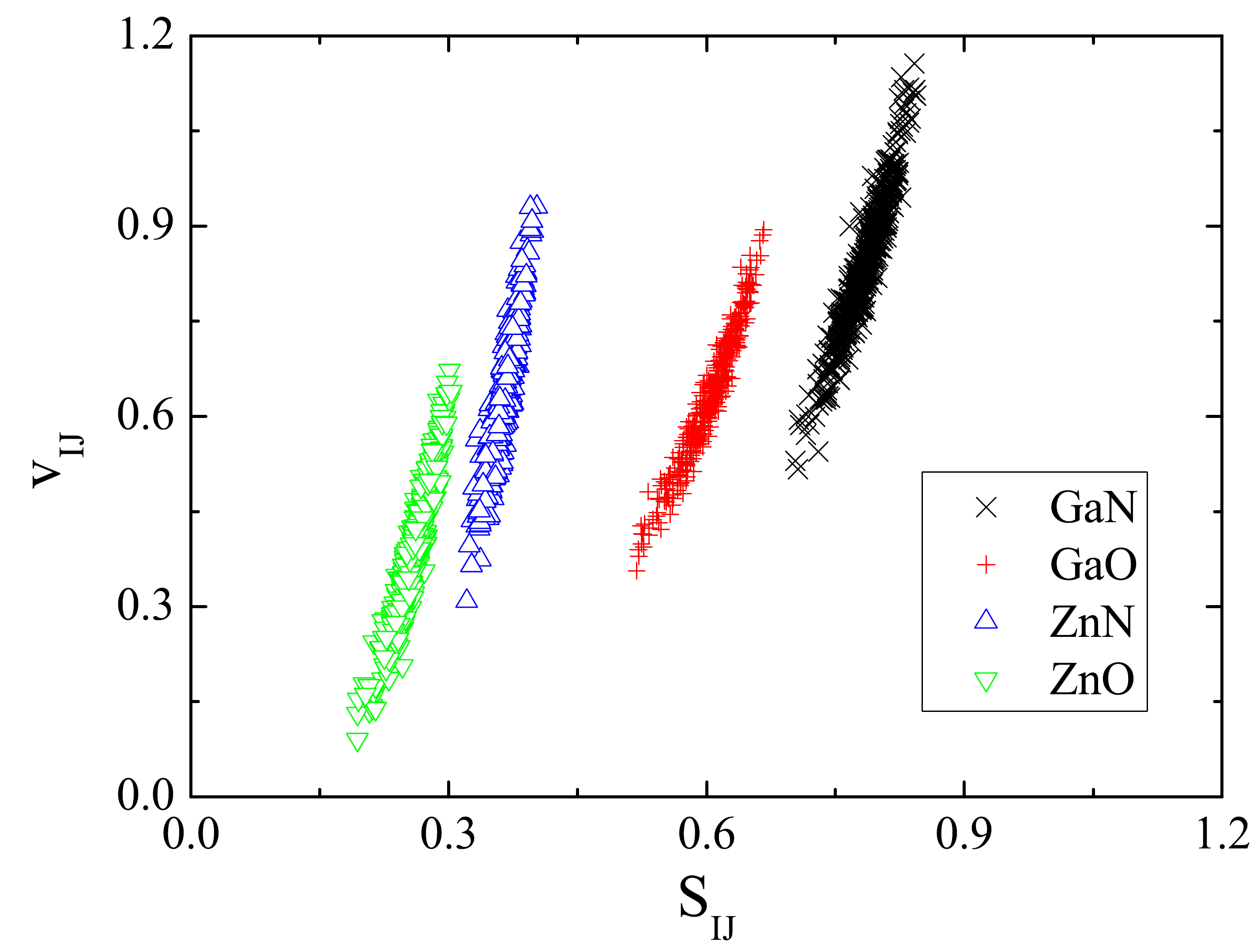}
\caption{The correlation $v_{IJ}\sim S_{IJ}$ for different types of bonds. $S_{IJ}$ is defined as $\sqrt{\sum_{i\in I, j\in J}S_{ij}^2}$.}
\end{figure}
\newline
The ability of BVM to predict the bond-length distribution relies significantly on the empirical correlation $v_{IJ}=\exp\left(\left(R_{IJ}^0-R_{IJ}\right)/b_{IJ}\right)$, the quality of which should be examined first. To yield accurate structural properties, DFT calculations are performed using the QUANTUM ESPRESSO package\cite{QE} with the PBEsol functional\cite{PBEsol}. The lattice constants of GaN and ZnO are well reproduced\cite{paper1}. In Table 1, the original (tabulated in Ref. \cite{BVM_para}) and fitted-to-DFT bond valence parameters are listed. As a sanity check, bond lengths of compound GaN and ZnO (labeled as $R^0$) calculated with the two sets of bond valence parameters are also listed. As the fitting procedure releases the freedom of the bond softness $b_{IJ}$, an overall improvement is observed for the fitted-to-DFT set of bond valence parameters. Fig. 4 shows the correlation between the DFT-calculated bond lengths and the BVM-predicted bond valences. Bond-length distribution is predicted by BVM with good accuracy. The prediction of bond-angle variation is less accurate, as shown in Fig. 5. The fitted bond valence parameters $k$ for Ga and Zn are 18.1$^\circ$/v.u. and 20.1$^\circ$/v.u. respectively. In order to perform pre-relaxation, one can simply add a penalty function to bond-length distribution and bond-angle variation. A large penalty to bond-length distribution is suggested while bond angles are subject to change.
\begin{table*}\small
\centering
\caption{\label{arttype} Bond valence parameters.}
\begin{tabular}{ccccccccccc}\hline
\br
&\multicolumn{1}{c}{}&\multicolumn{4}{c}{original BVM\cite{BVM_para}}&\multicolumn{1}{c}{}&\multicolumn{4}{c}{fitted to DFT}\\
&\multicolumn{1}{c}{}&GaN&GaO&ZnN&ZnO&\multicolumn{1}{c}{}&GaN&GaO&ZnN&ZnO\\
\mr
$R_{ij}^0 (\AA)$&\multicolumn{1}{c}{}&1.84&1.73&1.77&1.704&\multicolumn{1}{c}{}&1.844&1.755&1.831&1.756\\
$b_{ij} (\AA)$&\multicolumn{1}{c}{}&\multicolumn{4}{c}{0.37}&\multicolumn{1}{c}{}&0.357&0.391&0.268&0.312\\
\br
$R^0$ (\AA)&\multicolumn{1}{c}{}&1.946&--&--&1.960&\multicolumn{1}{c}{}&1.947&--&--&1.972\\
Expt. (\AA)\cite{property}&\multicolumn{1}{c}{}&1.95&--&--&1.977&\multicolumn{1}{c}{}&1.95&--&--&1.977\\
\br
\hline
\end{tabular}
\newline
\end{table*}
\begin{figure}[htb!]
\includegraphics[scale=0.28]{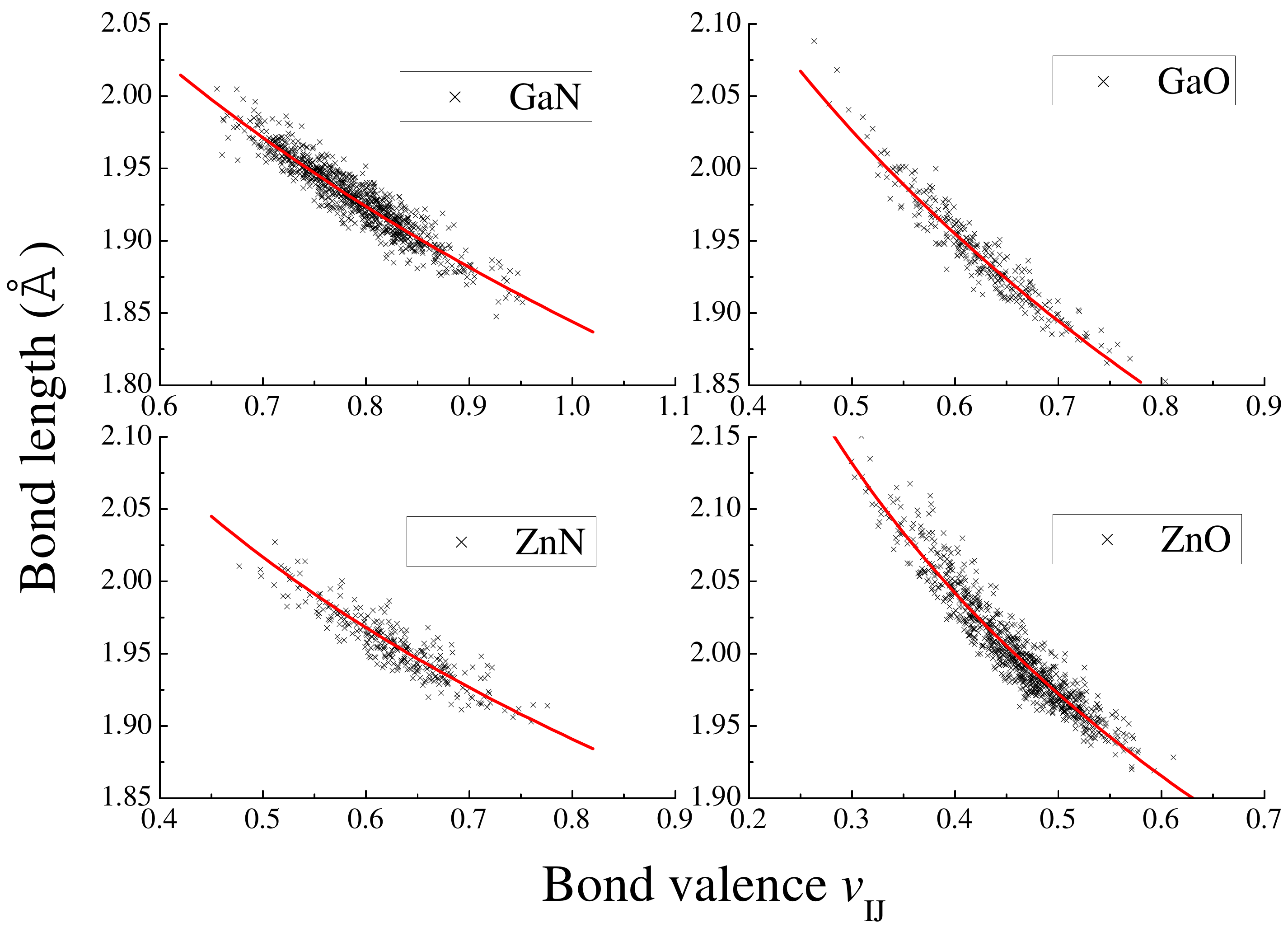}
\caption{Correlations between the DFT-calculated bond lengths and the BVM-predicted bond valences. The solid red lines represent the fitted correlations. In each figure the number of data points drawn is reduced by a factor of ten.}
\end{figure}
\vspace{-1pc}
\begin{figure}[htb!]
\includegraphics[scale=0.28]{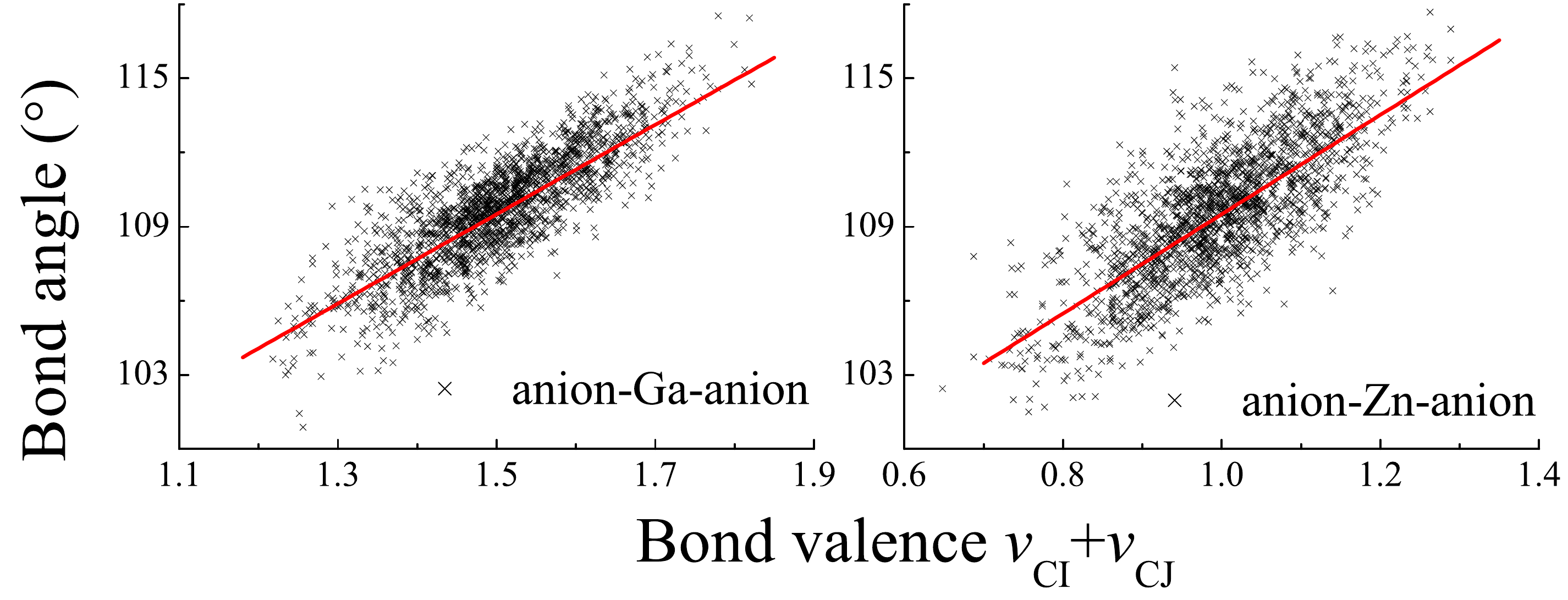}
\caption{Correlations between the DFT-calculated bond angles and the BVM-predicted bond valences. The solid red lines represent the fitted correlations. In each figure the number of data points drawn is reduced by a factor of ten.}
\end{figure}
\subsection{Bond-length distribution and bond-angle variation}
Ref. \cite{paper1} shows that there is strong SRO in the (GaN)$_{\rm{1-x}}$(ZnO)$_{\rm{x}}$ alloy. Although a completely random alloy may not be achievable under common experimental conditions, the degree of randomness introduced is influenced by the experimental methods adopted in growing the samples. For example, the kinetics of mixing may be facilitated by high-pressure\cite{exp2}. To represent the thermodynamic ensembles at $(x,T)$, Monte-Carlo simulations are performed on a DFT-based cluster expansion model. Given the site occupancies of each configuration, bond-length distribution and bond-angle variation are then predicted using the fitted-to-DFT bond valence parameters. The temperature measures the degree of randomness. Temperatures of 1123K, 2000K, 5000K and 20000K represent short-range ordered (SRO), disordered (DIS1 and DIS2) and random (RAN) alloy respectively. Fig. 6 shows the bond-length distributions at various temperatures. As the temperature is raised, the peak of bond-length distribution shifts slightly in the direction of shorter bond length. The shift of the peak position is small, and can be easily overwhelmed by other factors such as thermal expansion, which is not considered here. In the meanwhile the width of bond-length distribution becomes broader with increasing randomness. In Fig. 7, bond-length distributions of different types of bonds are shown. Upon mixing, the Ga-N bond shrinks while the Zn-O bond expands relative to the bond lengths in the corresponding compounds. From SRO alloy to RAN alloy, the shift is toward shorter bond length for Ga-N, barely temperature-dependent for Ga-O and Zn-N, and is reversed to the longer bond-length direction for Zn-O. This unusual tendency of bond-length distribution is a consequence of the non-isovalent nature of the alloy, and can be easily interpreted in terms of bond valence. One consequence of elevating the degree of randomness is to enhance the statistical presence of the energetically unfavored valence-mismatched Ga-O and Zn-N pairs. In the language of BVM, for a cation-anion pair, enhancing the presence of N(O) neighbors around the cation and Ga(Zn) neighbors around the anion will drain(pour) bond valence from(into) the cation-anion pair and as a result the bond is lengthened(shortened). Of particular importance is the Zn-N bond-length distribution due to the decisive role of Zn3$d$-N2$p$ repulsion on the top of the valence band. In Ref. \cite{paper1}, an almost linear band gap reduction upon increasing the ZnO content for the short-range ordered alloy is observed. Since the $p$-$d$ repulsion is inversely proportional to the bond length, upon increasing the ZnO content a shortened Zn-N bond-length distribution is expected, which is confirmed by the BVM prediction shown in Fig. 8. The stronger $p$-$d$ repulsion pushes the top of the valence band, resulting in the linear band gap reduction. Fig. 9 shows the anion-cation-anion bond-angle variation of short-range ordered alloy at $x=0.5$. The N-Ga-N angle expands while the O-Zn-O angle shrinks relative to the ideal tetrahedral angle 109.5$^{\circ}$, which can be explained by noticing that the bond valence of the ligand cation-O bond is generally smaller than that of the ligand cation-N bond. For Fig. 7-9, see Ref. \cite{paper1} for the DFT-calculated more reliable but less statistical predictions.
\begin{figure}[htb!]
\includegraphics[scale=0.28]{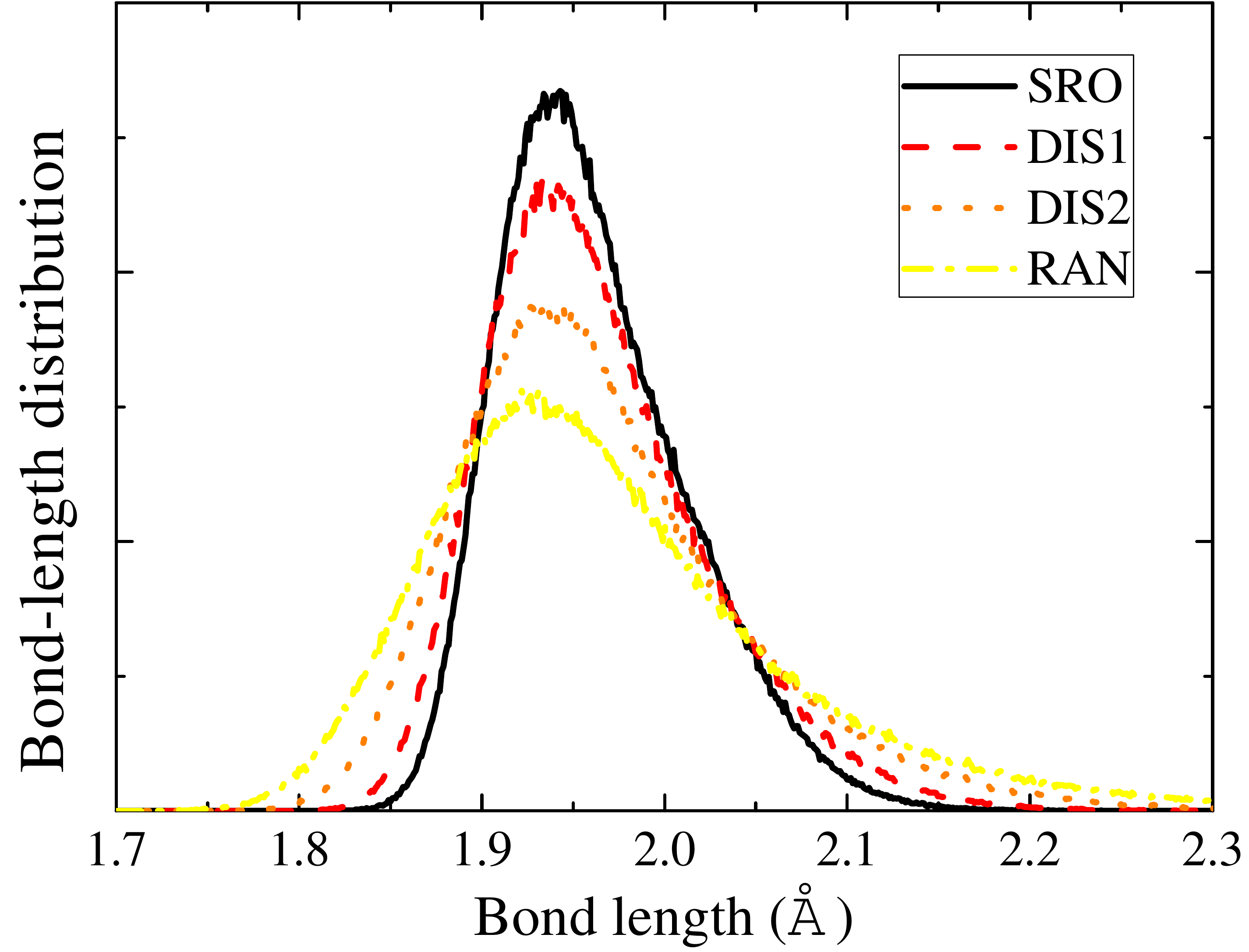}
\caption{Temperature dependence of bond-length distribution at $x=0.5$.}
\end{figure}
\begin{figure}[htb!]
\includegraphics[scale=0.28]{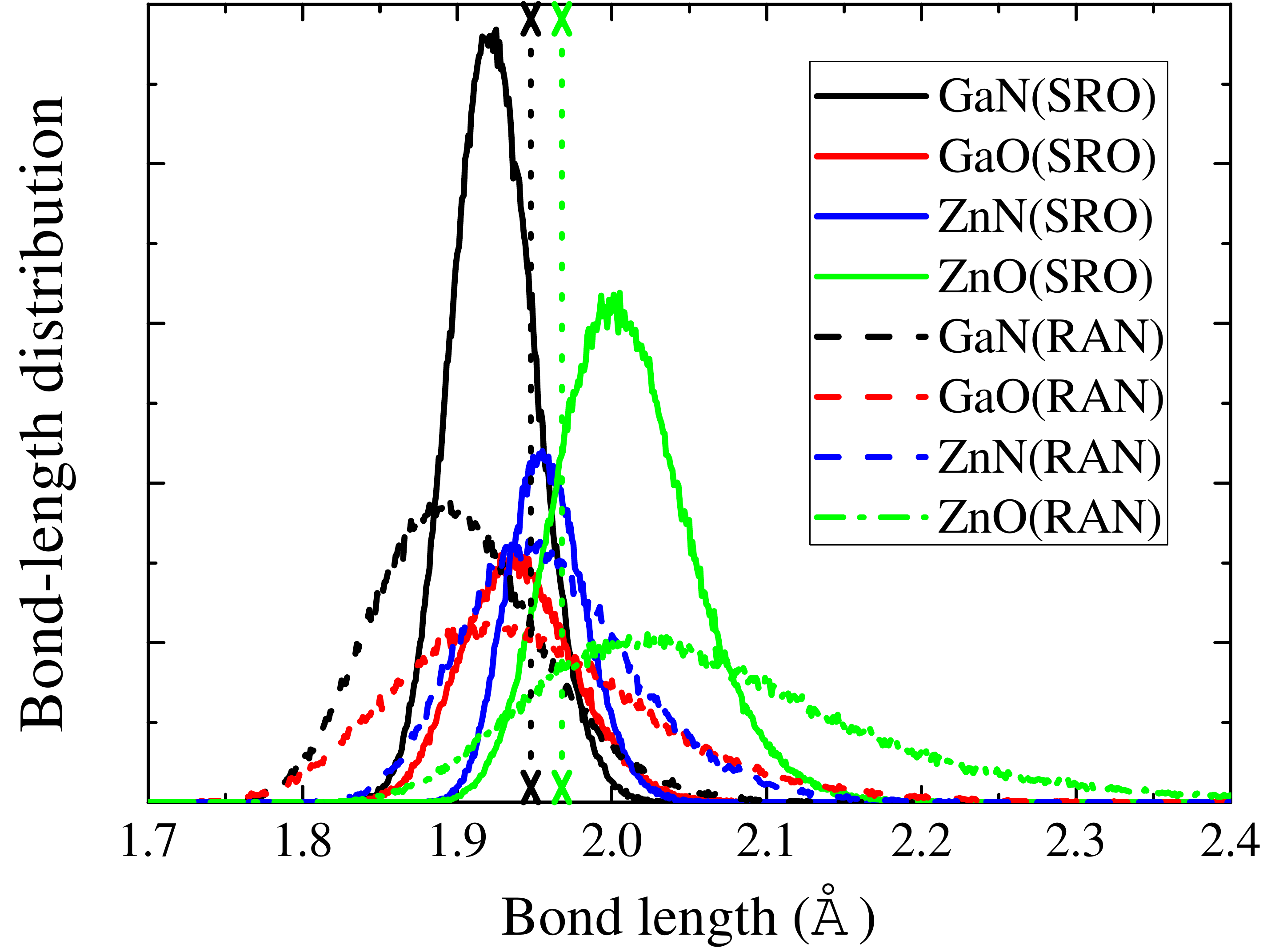}
\caption{Bond-length distributions of short-range ordered alloy and random alloy at $x=0.5$. The vertical dotted lines mark the bond lengths of the corresponding compounds.}
\end{figure}
\begin{figure}[htb!]
\includegraphics[scale=0.28]{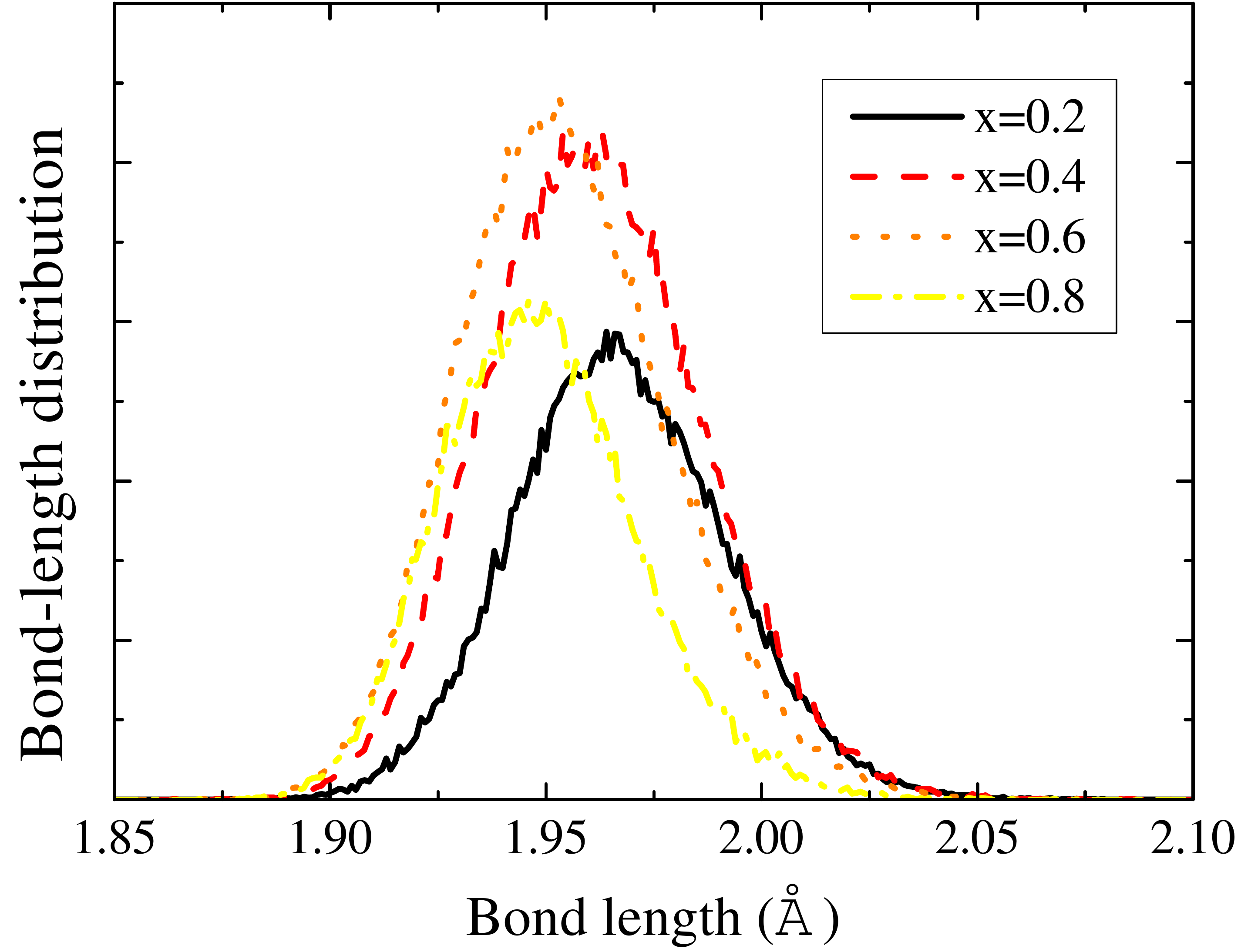}
\caption{Zn-N bond-length distribution at various ZnO content.}
\end{figure}
\begin{figure}[htb!]
\includegraphics[scale=0.28]{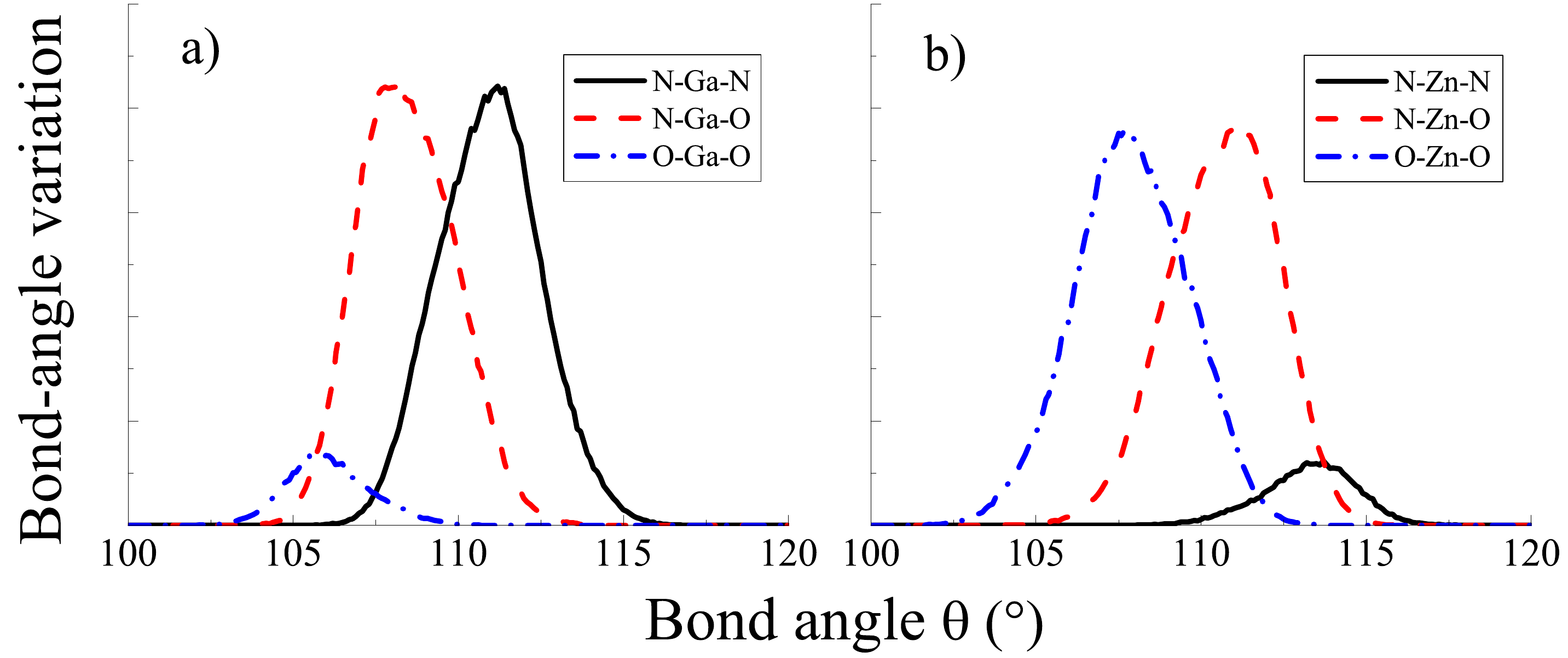}
\caption{The anion-cation-anion bond-angle variation of short-range ordered alloy at $x=0.5$.}
\end{figure}
\subsection{Energetics}
As for the energetics, DFT total energy calculations are performed on 170 structures selected from the $T=1,123$K thermodynamic ensemble over the full range of composition. The formation energies are also calculated using the $valence$ $loop$ $rule$ (i.e. the measure of energy $E=\alpha\sum_{I,J}v_{IJ}^2$). The results are shown in Fig. 10. The fitted parameter $\alpha=1.07$ is consistent with that of Ref. \cite{WangPRL}. The power of BVM is shown by the accurate reproduction of the energetics. However, BVM fails to reproduce the ordered superlattice (GaN)$_{1}$(ZnO)$_{1}$ ground state at $x=0.5$\cite{LL}, possibly due to the nearest-neighbor short-range nature of BVM itself, i.e. the wurtzite and zincblende structures are indistinguishable from one another in BVM. The formation energy of (GaN)$_{1}$(ZnO)$_{1}$ predicted by BVM is positive, while that predicted by DFT is slightly negative\cite{LL}. The discrepancy should not affect any conclusion drawn in present study since only the disordered phase is concerned.
\vspace{1pc}
\begin{figure}[htb!]
\includegraphics[scale=0.28]{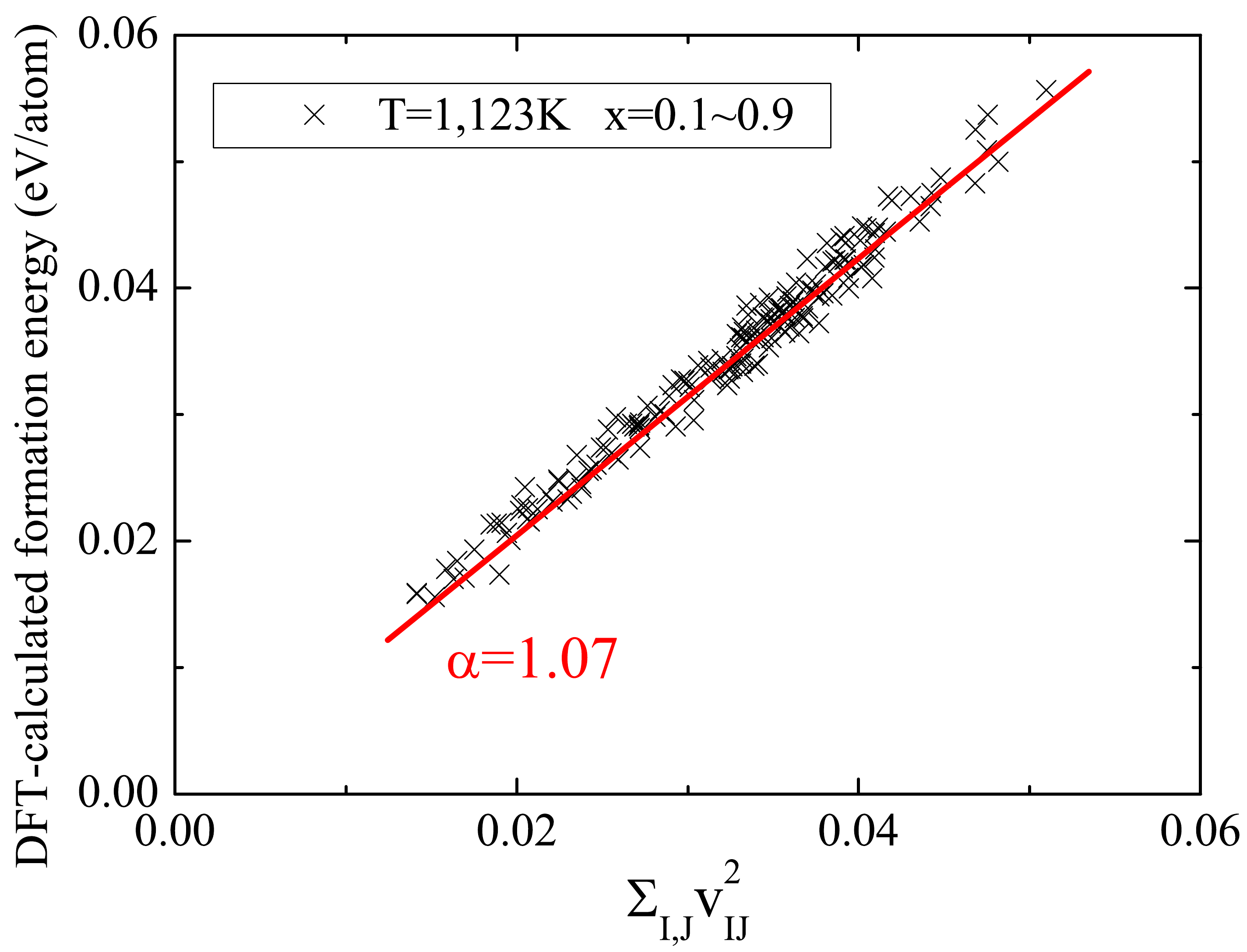}
\caption{Comparison of DFT-calculated formation energies with BVM-predicted formation energies.}
\end{figure}
\newline
Inclusion of vibrational entropy in the first-principles alloy phase diagram calculation is a long-standing challenge. The main difficulty lies in the conflict between the requirement for a large supercell and the expensive computational cost associated with it. The problem is partly alleviated by the SQS approach\cite{SQSphonon}\cite{SQSTFC}\cite{paper1}. Another idea is to use a bond-length-dependent transferable force constant\cite{Ceder}\cite{AlGaInN}, where the bond stiffness is predicted from the bond length and the chemical identity of the bond. The present study reveals the strong correlation between bond valence and bond length, which indicates the possibility of using bond valence instead of bond length as a predictor for bond stiffness. Such extension will release the estimation of nearest-neighbor force constants from the requirement for the knowledge of the relaxed geometry of a configuration. Fig. 11 shows the dependence of stretching bond stiffness $\phi_{\alpha\alpha}^{II}$ and $\phi_{\alpha\alpha}^{IJ}$ on bond length, where $\alpha$ refers to the bond-stretching direction and $I,J$ are nearest neighbors. The bond stiffness calculations are performed on selected 72-atom supercells with a displacement of $0.02\AA$ from the relaxed atomic coordinates along each bond direction. While a linear bond stiffness vs bond length relationship is suggested in bond-length-dependent transferable force constant approach\cite{Ceder}\cite{AlGaInN}, an exponential correlation (similar with that between bond valence and bond length) seems to fit better according to the present study, which is consistent with the interpretation that bond valence measures bond strength. Bond stiffness depends on bond length in a similar manner. The most covalent Ga-N bond is the stiffest, while the most ionic Zn-O bond is the softest.
\vspace{1pc}
\begin{figure}[htb!]
\includegraphics[scale=0.28]{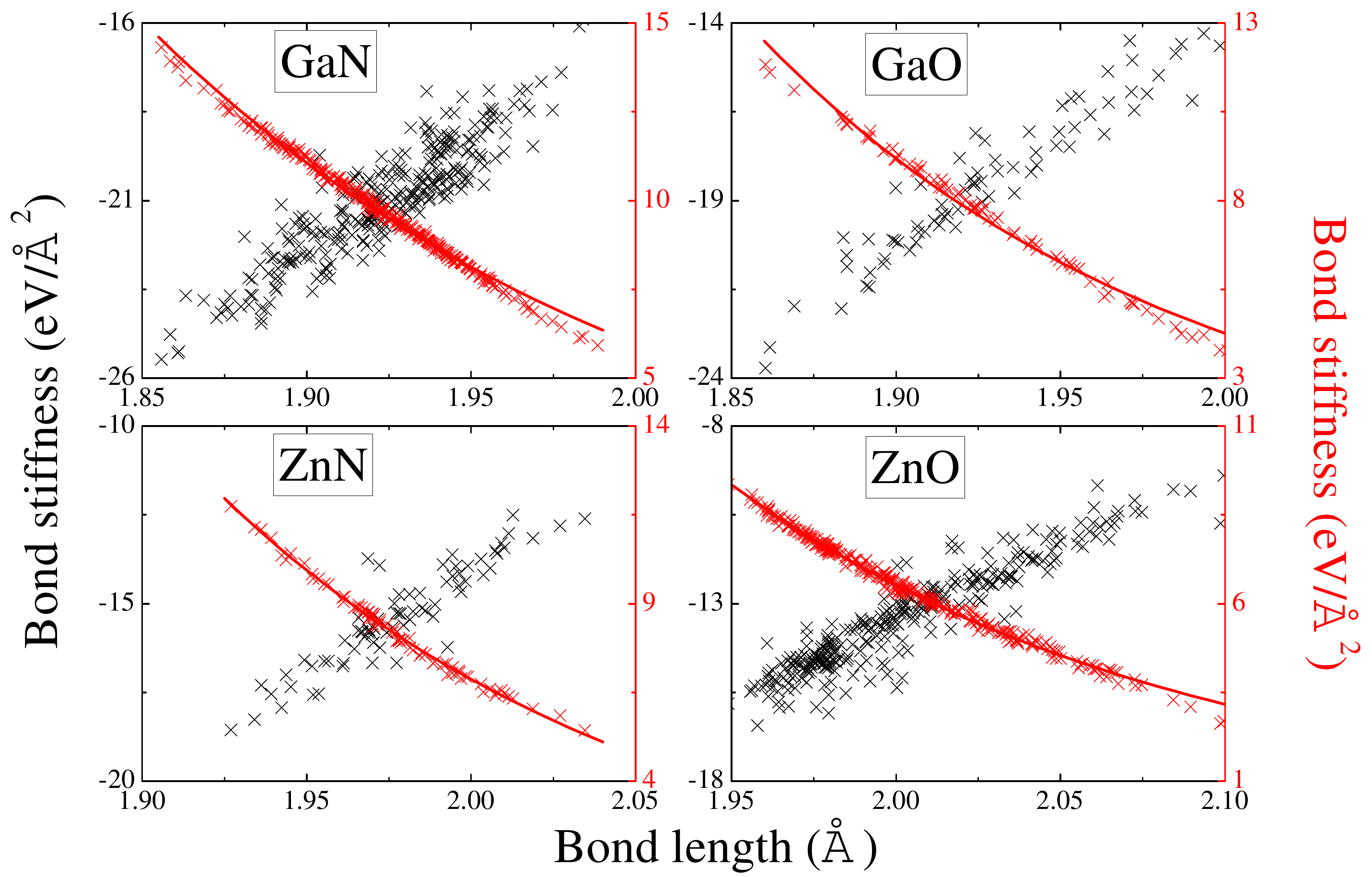}
\caption{Correlation between stretching bond stiffness and bond length. $\phi_{\alpha\alpha}^{II}$ is shown in black, and $\phi_{\alpha\alpha}^{IJ}$ is shown in red.}
\end{figure}
\newline
For isovalent \uppercase\expandafter{\romannumeral3}-\uppercase\expandafter{\romannumeral5} semiconductor alloys, the widely used Keating valence force field (KVFF) model\cite{Keating} yields generally good accuracy\cite{Zunger}\cite{Neugebauer}\cite{GaInNP}\cite{InGaN}\cite{Lopuszynski1}\cite{Lopuszynski2}. In KVFF, the force constants are related to the macroscopic elastic constants, and therefore can be accurately determined experimentally. Also the isovalent nature of \uppercase\expandafter{\romannumeral3}-\uppercase\expandafter{\romannumeral5} semiconductor alloys guarantees good transferability from compound semiconductors to the corresponding alloy. For non-isovalent semiconductor alloys, the transferability no longer holds, for the apparent reason that there exists no wurtzite GaO or ZnN. The present study offers an alternative way of accurately reproducing the energetics of non-isovalent semiconductor alloys with BVM, where only site occupancies are needed. The non-isovalent nature is well captured by the $valence$ $sum$ $rule$. An extension of the BVM leads to the modelling of an atomistic potential. In present study, the relaxation energy is assumed to consist of three parts:
\begin{equation}
\eqalign{E_{relax}=&\alpha\sum_{I,J}v_{IJ}^2+\sum_{\rm{I=Ga,Zn}}k_I\sum_{J_1,J_2}\left(\theta_{J_1IJ_2}-\theta_0\right)^2\cr
&+\sum_{\rm{I=N,O}}\beta_I\left(\sum_{J}v_{IJ}-V_{0,I}\right)^2}
\end{equation}
The first term is simply the $valence$ $loop$ $rule$, and the second term is the harmonic angle potential. The third term accounts for large lattice relaxations by penalizing deviations from the bond valence conservation and is important for reliable molecular dynamics simulations\cite{perovskite1}\cite{perovskite2}. In the fitting procedure each relaxed structure is expanded and contracted by 1\%. Fitting parameters $k_{\rm{N,O}}$ and $\beta_{\rm{Ga,Zn}}$ are found to be negligible. In Figure 12 the comparison between DFT-calculated and BVM-fitted formation energies is shown. The agreement is generally satisfactory. Further studies will involve refinement of the atomic potential.
\begin{figure}[htb!]
\includegraphics[scale=0.28]{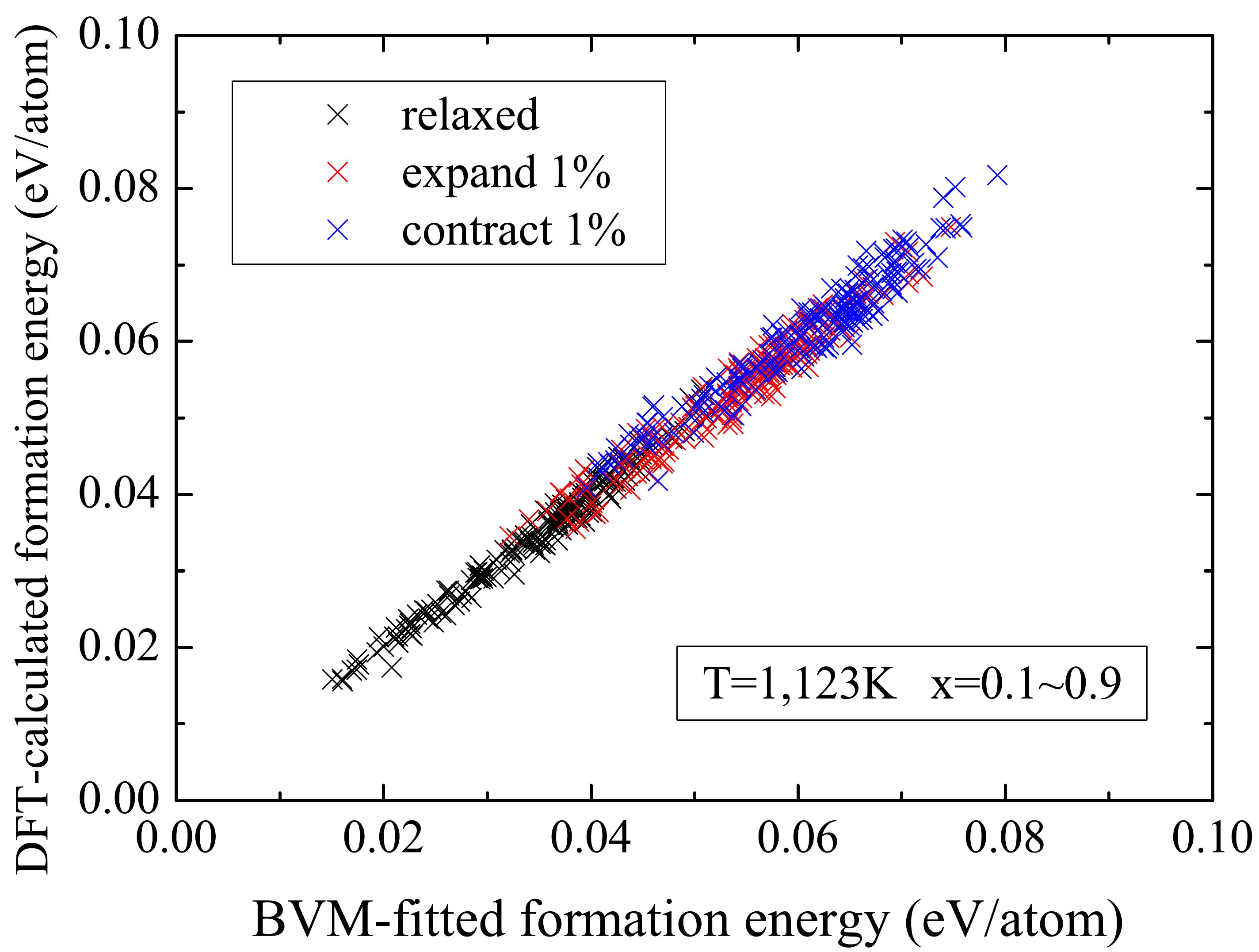}
\caption{Comparison between DFT-calculated and BVM-fitted formation energies.}
\end{figure}
\section{Discussion and Conclusions}
A physical interpretation of BVM is discussed from the computational perspective. The underlying assumptions and correlations within BVM are revealed by DFT calculations on the non-isovalent semiconductor alloy (GaN)$_{\rm{1-x}}$(ZnO)$_{\rm{x}}$. Bond-length distribution and bond-angle variation are predicted by fitting BVM empirical relations to reliable DFT-calculated structural data. The unusual relaxations associated with the non-isovalent nature of the alloy are explained. Effects of SRO on bond-length distribution and bond-angle variation are also discussed. The energetics is accurately reproduced by BVM. The connection between bond valence and stretching bond-length-dependent transferable force constant is revealed. A tentative improved bond valence potential is proposed. In principle, the methods of the present study should also be applicable for other non-isovalent semiconductor alloys.

\ack
JL thanks Philip B. Allen and Maria V. Fern\'{a}ndez-Serra for fruitful discussions. This research used computational resources at the Center for Functional Nanomaterials, Brookhaven National Laboratory, which is supported by the US Department of Energy under Contract No. DE-AC02-98CH10886. Work at Stony Brook was supported by US DOE Grant No. DE-FG02-08ER46550. JL is sponsored by China Scholarship Council.

\section*{References}
\bibliographystyle{iopart-num}
\bibliography{ref}

\end{document}